\documentclass[notitlepage,superscriptaddress,prl,reprint]{revtex4-1}
\usepackage{url,graphicx,braket,mathdots}
\begin{document}

\def\Gv{\mathbf{G}}
\def\Iv{\mathbf{I}}
\def\Sv{\mathbf{S}}
\def\Fv{\mathbf{F}}
\def\Nv{\mathbf{N}}
\def\Rv{\mathbf{R}}
\def\Jv{\mathbf{J}}
\def\Lv{\mathbf{L}}
\newcommand{\Dstate}[0]{\ensuremath{A'^{2}\Delta_{3/2}}}
\newcommand{\Pstate}[1]{$A^{2}\Pi_{#1/2}$} 
\newcommand{\Sstate}[0]{\ensuremath{X^{2}\Sigma}}

\title{3-D Magneto-Optical Trap of Yttrium Monoxide}
\author{Alejandra L. Collopy}
\affiliation{JILA, National Institute of Standards and Technology and Department of Physics, University of Colorado, Boulder, CO 80309, USA}
\author{Shiqian Ding}
\affiliation{JILA, National Institute of Standards and Technology and Department of Physics, University of Colorado, Boulder, CO 80309, USA}
\author{Yewei Wu}
\affiliation{JILA, National Institute of Standards and Technology and Department of Physics, University of Colorado, Boulder, CO 80309, USA}
\author{Ian A. Finneran}
\affiliation{JILA, National Institute of Standards and Technology and Department of Physics, University of Colorado, Boulder, CO 80309, USA}
\author{Lo\"{i}c Anderegg}
\affiliation{Department of Physics and Center for Ultracold Atoms, Harvard University, Cambridge, MA 02138, USA}
\author{Benjamin L. Augenbraun}
\affiliation{Department of Physics and Center for Ultracold Atoms, Harvard University, Cambridge, MA 02138, USA}
\author{John M. Doyle}
\affiliation{Department of Physics and Center for Ultracold Atoms, Harvard University, Cambridge, MA 02138, USA}
\author{Jun Ye}
\affiliation{JILA, National Institute of Standards and Technology and Department of Physics, University of Colorado, Boulder, CO 80309, USA}

\begin{abstract}
We report three-dimensional trapping of an oxide molecule (YO), using a radio-frequency magneto-optical trap (MOT). The total number of molecules loaded is $\sim$1.5$\times10^4$ , with a temperature of 7(1)~mK. This diversifies the frontier of molecules that are laser coolable and paves the way for the second-stage narrow-line cooling in this molecule to the microkelvin regime. Futhermore, the new challenges of creating a 3-D MOT of YO resolved here indicate that MOTs of more complex non-linear molecules should be feasible as well.
\end{abstract}


\maketitle

The last decade has seen considerable successes in the association of atomic quantum gasses to produce ultra-cold molecules \cite{Moses} and in sympathetic cooling of molecular ions with laser coolable atomic species \cite{Hudson, Chou}. But, only very recently has laser cooling of neutral molecules directly from a beam to the millikelvin regime and below become viable. Other promising techniques for diverse molecule cooling include electro-optic cooling, which has been demonstrated for formaldehyde \cite{Zeppenfeld}, and various deceleration techniques, such as Stark \cite{StuhlOH} or centrifugal deceleration \cite{Wu}. Molecular laser cooling is expanding the number of species usable for cold molecule experiments where large dipole moments and strong intermolecular interactions exist, as well as enlarging the workspace for cold chemistry studies \cite{Bohn, Carr}.  

Laser cooling is only logistically feasible for a subset of molecules with sufficiently closed cycling transitions \cite{DiRosa}. The first proposal for a molecular magneto-optical trap (MOT) identified five classes of diatomic molecules as being amenable to laser cooling and trapping \cite{Stuhl}, including halides, hydrides, oxides, carbides, and sulfides. However, only the fluorides SrF \cite{Barry2014} and CaF \cite{Truppe, Williams, Anderegg} have thus far been loaded into 3-D MOTs. Since the initial proposals for diatomic species, a large number of polyatomic pseudofluorides have also been identified as laser coolable \cite{Kozy2016MOR, Isaev2016Poly} and this has been demonstrated with SrOH \cite{Kozy2017SrOH}. Additionally, even in the absence of laser cooling or trapping, quasi-closed cycling transitions that enable rapid photon cyling can be utilized for internal state preparation and efficient readout in precision measurement experiments such as searches for the electron electric dipole moment (EDM) \cite{Lim, Roussy, Panda}, or parity violation (PV)\cite{Altuntas}. Furthermore, extension of EDM and PV experiments to diatomic and polyatomic molecules that can be directly cooled in a MOT promises great advances in sensitivity, due to vastly increased coherence times and the potential for suppression of systematic errors \cite{Lim, Kozy2017EDM, Isaev2017RaOH}. However, as laser cooling and trapping is extended to more diverse diatomic and polyatomic species, new complications in molecular structure will arise beyond those present in the alkaline-earth monofluorides studied to date. Our work with yttrium monoxide thus represents a crucial milestone in extending 3-D MOTs beyond ``simple'' alkaline-earth diatomic molecules like SrF and CaF to more complex molecular species desirable for many applications in quantum science and technology \cite{Wall2013, Wall2015}.

Building on our earlier work of two-dimensional (2-D) magneto-optical trapping \cite{Hummon} and longitudinal slowing of yttrium monoxide (YO) \cite{Yeo}, in this Letter we report a 3-D MOT of YO. This is the first realization of a MOT for an oxide molecule. The presence of an intermediate electronic state in the YO molecule also represents a new intricacy that is now overcome in the molecular cooling field. In this case, the particular choice of repumping scheme becomes more important to maintain a quasi-closed cycling transition and a large optical scattering rate. The ground-to-intermediate state transition in YO is also amenable to further laser-cooling, making this MOT instantiation a critical step towards the realization of narrow-line cooling in a molecule. As narrow-line cooling greatly expanded the physics available for alkaline-earth \cite{Katori, Loftus} and magnetic atoms \cite{Frisch}, it too may yield rich fields of study for molecules.

\begin{figure}[]
    \includegraphics[width=.5\textwidth]{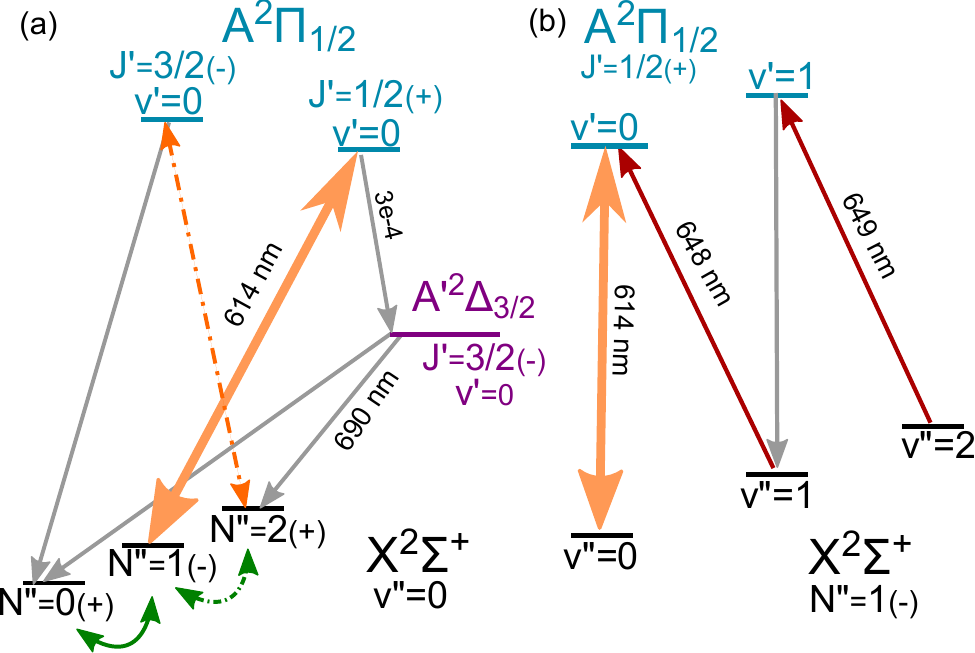}

  \caption{Relevant level structure of the YO molecule. (a) The solid orange double-headed arrow indicates the main cooling transition, while green curved double-headed arrows depict microwave mixing. Gray arrows are decay channels. Dashed-dotted lines indicate various options we have for repumping states. (b) Red arrows indicate vibrational repump lasers which are used to limit unrecovered vibrational branching to the 10$^{-6}$ level.}
  \label{fig:levels} 
\end{figure}

\begin{figure}[]
\includegraphics[width=.5\textwidth]{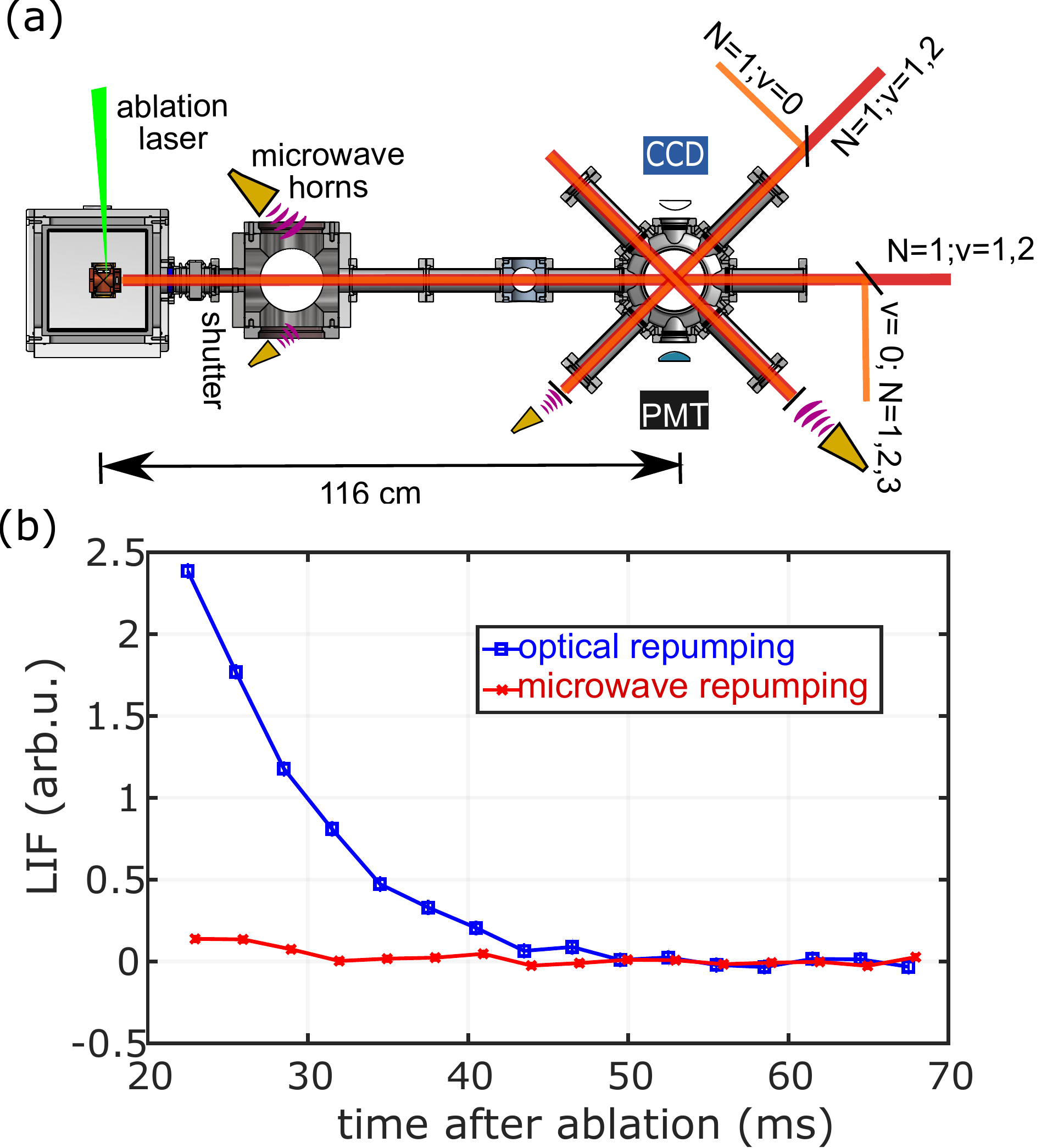}
\caption{(a) Top view of experimental apparatus. Molecules are produced at the left and travel to the right while interacting with a counterpropagating slowing beam. Slowed molecules are loaded into the MOT at the right, where they are imaged. Microwave horns are coupled into the chamber along the slowing region and in the MOT region. (b) Velocity sensitive detection of molecules at 5~m/s after the slowing sequence. The blue (square) line utilized optical repumping of \textbf{N}$^{\prime\prime}$=2 molecules, whereas the red (x) line used microwave repumping of these molecules. Error bars are within symbol size. }
\label{fig:schematic}
\end{figure}

The level structure of YO has been described in previous work \cite{Hummon,Yeo,Collopy}, and so here we highlight only the most salient details. In order to achieve a nearly closed optical cycling transition in YO, we utilize one of the two schemes depicted in Fig.~\ref{fig:levels}(a). Two repump lasers limit vibrational branching loss to the  10$^{-6}$ level. By using the \textbf{N}$^{\prime\prime}$=1$\rightarrow$ \textbf{J}$^\prime$=1/2 transition, parity and angular momentum selection rules ensure rotational closure. However, because of leakage to the intermediate \Dstate{}~state, molecules will decay to \Sstate{}(\textbf{N}$^{\prime\prime}$=0,2) after $\sim$3000 photon-scattering events \cite{Yeo}. Although the transition from the \Dstate{}~state to the \Sstate{}~state is nominally forbidden, we find it does occur due to mixing of the \Dstate{}~state with the nearby \Pstate{3}~state, giving the \Dstate{}~state a lifetime of $\sim$1~$\mu$s \cite{Collopy}. We note that the transition wavelength to directly repump the \Dstate~state to the \Pstate{1}(\textbf{J}=3/2)~state is 5.6~$\mu$m and would be experimentally difficult to implement and frequency stabilize. To repump from \Sstate{}(\textbf{N}$^{\prime\prime}$=0), we apply microwaves at 23~GHz to remix the \Sstate{}(\textbf{N}$^{\prime\prime}$=0,1) levels. The microwaves couple states of same quantum number \textbf{G}=\textbf{I}+\textbf{S}, and are modulated to cover the necessary hyperfine manifolds within \textbf{G}$=1$. To recycle the population in \Sstate(\textbf{N}$^{\prime\prime}$=2), we either use an optical repump to the \Pstate{1}(\textbf{J}$^\prime$=3/2) level or use microwaves to directly remix the \Sstate{}(\textbf{N}$^{\prime\prime}$=1,2) levels, split by 46~GHz. Due to the number of states involved in the cyling transition, the optical repumper increases the scattering rate as opposed to microwave mixing by decoupling the \Sstate{}(\textbf{N}$^{\prime\prime}$=2) level from the main cycling transition. The change realized in this manner increases the maximum possible scattering rate from $\Gamma$/13 to $\Gamma$/8, a gain of a factor of 1.6, where $\Gamma$=$2\pi\times5$~MHz is the natural linewidth of the \Pstate{1}~state. In practice, with optical repumping of the \Sstate{}(\textbf{N}$^{\prime\prime}$=2) level, we realize a scattering rate of $\sim$10$^6$ s$^{-1}$, which is measured by removing each vibrational repump and observing the rate of decay into a dark state in comparison with known Franck-Condon branching ratios, akin to the method used in \cite{Yeo}. Molecules which start in or find themselves in the \Sstate{}(\textbf{N}$^{\prime\prime}$=3) level due to off-resonant excitation to unintended excited states are repumped optically through the \Pstate{1}(\textbf{J}$^\prime$=5/2, v$^\prime$=0) state. 

\begin{figure*}[]
\includegraphics[width=\textwidth]{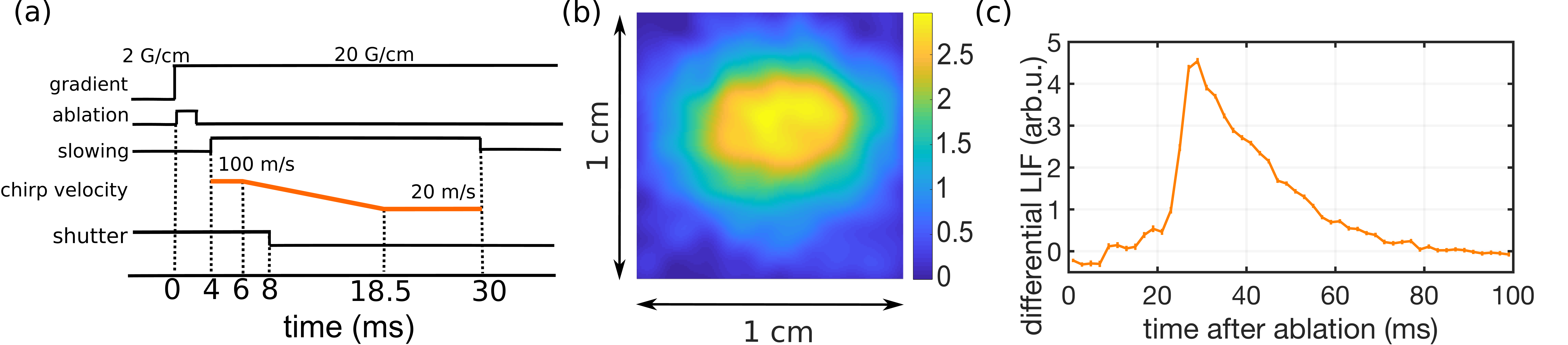}
\caption{\label{fig:mot}(a) Timing sequence used for MOT production. MOT laser beams are on at all times. (b) Phase-subtracted CCD image of the MOT, with a Gaussian filter of $\sigma=0.5$~mm. (c) Phase-subtracted LIF time trace of MOT loading and decay acquired from the PMT.}
\end{figure*}

Although the \Sstate{}(\textbf{N}$^{\prime\prime}$=1)$\rightarrow$\Pstate{1}(\textbf{J}$^\prime$=1/2) transition prevents rotational branching, it requires that we excite from a greater \textbf{F}$^{\prime\prime}$ to a smaller \textbf{F}$^\prime$. This is often referred to as a type II cooling transition, as opposed to the type I transition more typically seen in atomic laser cooling experiments. In order to prevent decays into a dark m$_F$ polarization state during slowing or trapping, it is necesssary to utilize a remixing method. We modulate the polarization of our laser light between orthogonal linear (circular) polarization while slowing(trapping) at a rate of about 5~MHz. This rate was chosen because in our 1-D cooling experiments the transverse laser-cooled temperature achieved continued to decrease up to that switching rate \cite{Hummon}. The polarization switching also necessitates care for the choice of magnetic fields during trapping, so as not to spend any time in an anti-trapping configuration.  As implemented previously for YO in one and two dimensions \cite{Hummon}, we continue to use an AC MOT in this 3-D implementation. The polarization switching of the MOT beams is implemented with a Pockels cell to change between $\sigma^+$ and $\sigma^-$ polarization synchronously with the magnetic field oscillation. While DC MOTs have been made for molecules, the effectiveness of this method relies on the small but non-zero magnetic g-factor in the excited \Pstate{1} state \cite{Tarbutt} or a bichromatic scheme \cite{Truppe, TarbuttPRA}. Comparisons between AC and DC MOTs for molecules have shown that the AC version provides a slightly better performance \cite{Anderegg,Norrgard}.

The apparatus used to produce, decelerate, and trap the molecules is depicted in Fig. \ref{fig:schematic}(a).
In contrast to previous work on slowing of YO \cite{Yeo}, we utilize a single stage cryogenic buffer gas cell as opposed to a two-stage cell. The single stage cell has larger peak molecule velocity (120 m/s instead of 70 m/s), but higher brightness. The molecules are produced via ablation of a Y$_2$O$_3$ ceramic target with a doubled (532~nm) Nd:YAG 7~mJ pulse.  In order to improve the performance of the cell at low helium buffer gas flow rates, we find that coating the interior of the cell with a layer of Y$_2$O$_3$ nanopowder yields a larger flux of molecules for a given helium buffer gas flow rate. The nanopowder coating also brings the helium buffer gas flow rate that produces the maximal number of YO molecules from 8 to 2~sccm. The slowing laser beam is counterpropagating to the molecular beam. An in-vacuum shutter positioned 28~cm downstream from the buffer gas cell cuts off the continuous 0.5 sccm He flow after the YO molecule pulse has passed as well as preventing heating of the buffer gas cell due to the power in the slowing laser beam. 

Our slowing scheme is similar to previous work \cite{Yeo}, except that now, for simplicity, we no longer apply white light modulation to the slowing beam. We maintain use of a frequency chirp, with variable duration and chirp range of 100 to 20~m/s, to continually address the molecules as they slow down. Microwaves are coupled into the chamber partway down the molecule slowing distance and into the MOT region with microwave horns. Molecules in the trap region are imaged with a charge-coupled device (CCD) camera for spatially resolved imaging, and with a photo-multiplier tube (PMT) for time resolved information. We detect the laser induced fluorescence (LIF) of the molecules on the main cycling transition at 614~nm. We estimate our imaging photon collection efficiency to be about 0.3~\% and 0.1~\% for the CCD and PMT systems, respectively. 

 By frequency chirping the  main cooling laser and repump lasers in the counterpropagating beam, we slow  molecules from production speeds of approximately 120~m/s to trappable velocities of less than 5~m/s. We confirm the slowing by detecting the molecular LIF with a low-power Doppler sensitive probe beam at 45 degrees with respect to the molecular beam propagation, as shown in Fig. \ref{fig:schematic}(b). By changing from microwave to optical repumping of the \Sstate{}(\textbf{N}$^{\prime\prime}$=2) level (and a commensurate change in frequency chirp rate), we realize a gain in slowed molecule signal of a factor of 19. We estimate the total number of molecules slowed to 5~m/s using the optical repumping scheme of  \Sstate{}(\textbf{N}$^{\prime\prime}$=2) to be about 1$\times10^{5}$. When utilizing a higher scattering rate, we suffer less from the transverse divergence of the molecular beam due to less time required to reach the trapping region.
 
After slowing, molecules are loaded into the 3-D MOT.  The AC MOT coils in our experiment are an in-vacuum resonant circuit tunable from 3-5~MHz. By driving the circuit with 60~Watts (2~Amps through each coil), we achieve a RMS field gradient of 20~G/cm.  A timing diagram used for loading molecules is shown in Fig.~\ref{fig:mot}(a).  We note that the slowing beam stays on during the loading of the MOT; its endpoint of 20 m/s is equivalent to 32~MHz detuning, and so serves to gently provide the last bit of slowing to the very slow molecules. We find maximal loading with a field gradient of 20~G/cm and a red detuning of 9~MHz of all hyperfine components from resonance of the v$^{\prime\prime}$=0 laser light in the MOT beams. Vibrational repumps are spatially overlapped with MOT beams, while \Sstate{}(\textbf{N}$^{\prime\prime}$=2,3) are repumped along the slowing axis. In order to distinguish untrapped molecules still transiting the MOT region from the MOT itself, we examine a phase-subtracted differential signal of the correct phase minus 180 degrees out of phase between the MOT coil drive and the polarization switching of the MOT beams. 

The retroreflected MOT beams have a 1/e$^2$ beam waist of 10~mm, and have 25~mW per beam of v$^{\prime\prime}$=0 light. Repump powers are typically 25 and 8~mW per MOT beam for v$^{\prime\prime}$=1 and 2 respectively, with 32 and 8~mW for \textbf{N}$^{\prime\prime}$=2 and 3 (v$^{\prime\prime}$=0) respectively. The resulting MOT image after loading with a 20~G/cm RMS field gradient with 6~MHz red detuning of all v$^{\prime\prime}$=0 components is shown in Fig.~\ref{fig:mot}(b). The peak number loaded into the trap is approximately $1.5\times 10^4$ molecules (within a factor of 2), which we determine from our collection efficiency and peak photon scattering rate.

At normal MOT beam intensity, the 1/e lifetime of the MOT (as shown in Fig. \ref{fig:mot}(c)) is currently limited to $\sim$20~ms by an as-yet undetermined loss process. Optical repumping of the \Sstate{}(\textbf{N}$^{\prime\prime}$=3) level yields 30 percent more molecules in the MOT, but has no influence on the observed lifetime. We have also experimented with optical repumping of the \Sstate{}(v$^{\prime\prime}$=3) state via the \Pstate{1}(v$^\prime$=2) state, which has not yet shown an effect in our experiment (consistent with the expected Franck-Condon factors for YO and the number of photons scattered). We suspected that the largest leak from our quasi-closed optical cycle was via the \Dstate{} state to the \Sstate{}(v$^{\prime\prime}$=1, \textbf{N}$^{\prime\prime}$=0,2) levels, but found that applying microwaves to remix those states did not increase the lifetime, implying the loss is due to a different mechanism. We do find that the lifetime of the MOT increases when the MOT beam intensity is lowered, at the expense of loaded molecule number.

\begin{figure}[]
\includegraphics[width=0.5\textwidth]{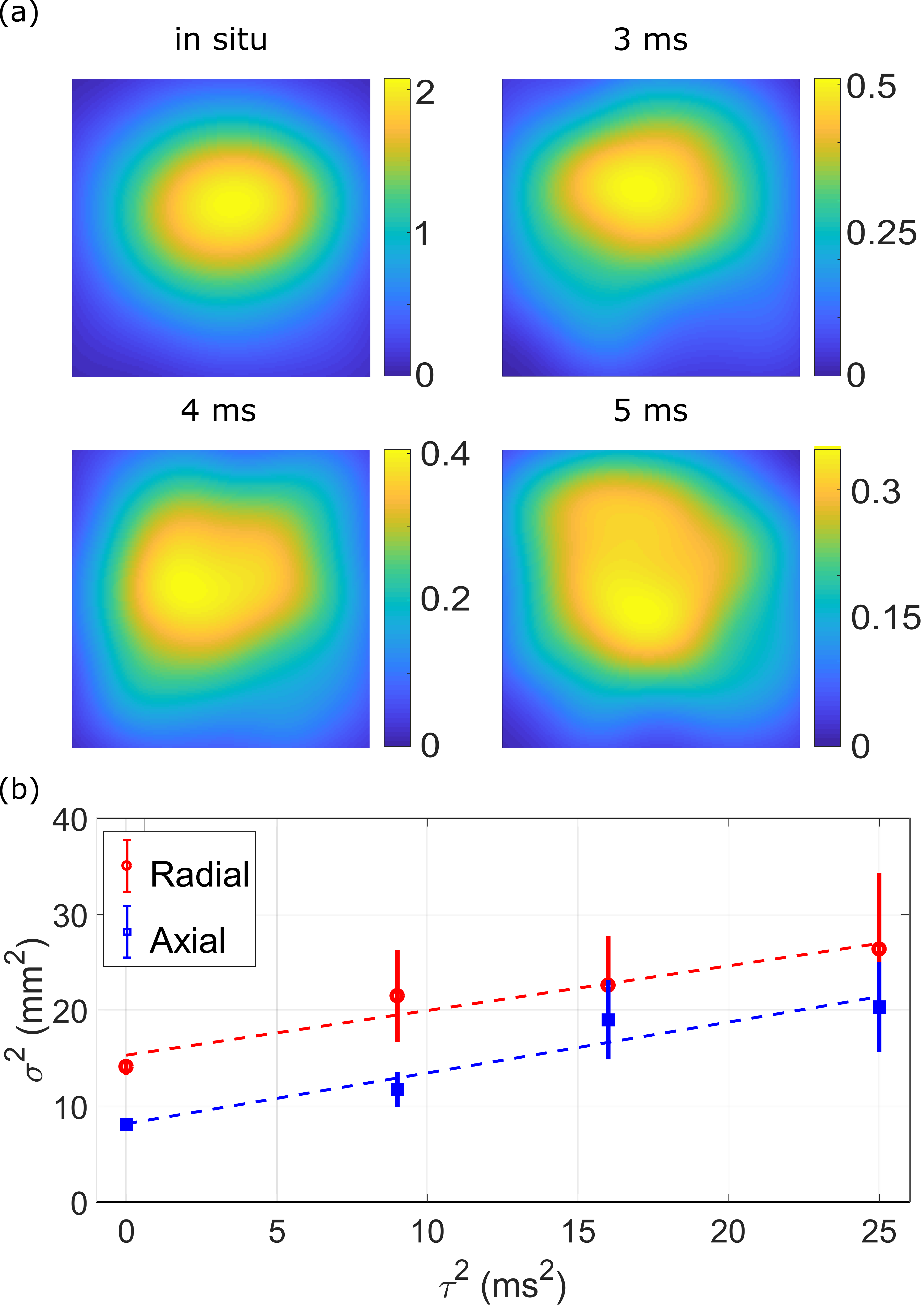}
\caption{MOT temperature characterization. (a) Phase-subtracted imaging  of the MOT after some period of free expansion. The CCD exposure time is 2~ms for each image, and is the average of 180 experimental repetitions with a Gaussian filter of $\sigma=1.5$~mm. (b) The measurement of the Gaussian width $\sigma$ of the cloud in the axial and radial directions as a function of free flight time $\tau$ determines the temperature in these two directions.}
\label{fig:temperature}
\end{figure}

To characterize the temperature of the MOT, as shown in Fig. \ref{fig:temperature}, we load molecules into the MOT at 20~G/cm and 6~MHz red detuning, and then switch off the MOT beams to allow free expansion for a variable amount of time. By fitting the cloud to a Gaussian distribution and measuring the width as a function of free-flight time, we extract the radial and axial temperature of the MOT to be $7(1)$~mK and $6.4(9)$~mK, respectively. 
Based on the size and temperature of the molecular cloud, we apply the equipartition theorem to determine the radial and axial trap frequencies to be  $f_{\text{radial}}$=30~Hz and $f_{\text{axial}}$=40~Hz.

In conclusion, we demonstrate the three dimensional magneto-optical trapping of YO. To our knowledge, this represents the first MOT of a non-fluoride molecule, broadening the horizon of chemical diversity of cooled molecules to include oxides, which are of chemical \cite{Chalek}, atmospheric, and astrophysical \cite{Bernard} relevance. It also represents the first MOT for a molecule with an electronic state intermediate to the main cycling transition states, which necessitates careful consideration of repump choice in order to rapidly cycle photons wihout losing molecules to dark states. Optical cycling in the presence of intermediate electronic states as well as decay to dark excited rotational levels both arise naturally when one is trying to create a MOT of non-linear polyatomic molecules \cite{Kozy2016MOR}. Our work with YO demonstrates that both of these issues can be efficiently resolved, opening the prospects of 3-D MOTs for complex non-linear molecules which have been proposed for a new generation of precision measurement \cite{Kozy2017EDM} and quantum simulation experiments \cite{Wall2015Chapter}.

This realization of a MOT of YO is a necessary step for further narrow-line cooling on the 150~kHz transition from the \Sstate~state to the \Dstate~state at 690~nm as proposed in Ref. \cite{Collopy}. Narrow-line cooling on this transition will allow direct cooling to the microkelvin regime, allowing the molecules to be readily loaded into magnetic or optical dipole traps~\cite{2008sawyer,Mccarron2017,Anderegg2018,Williams}. The transition has already been observed and characterized. Cooling on this transition will be experimentally more straightforward to implement than the initial trap due to a fewer number of photon scatterings needed for cooling and the ability to run in a type I MOT configuration at the expense of a reduced photon scattering rate. We also note that the finite lifetime of the first-stage broad linewidth MOT should be sufficient for loading to the second stage narrow-line MOT.

\begin{acknowledgements}
We thank Ivan Kozyryev for feedback on the manuscript. We acknowledge funding support from ARO-MURI, Gordon and Betty Moore Foundation, NIST, and the NSF Physics Frontier Center. 
\end{acknowledgements}

\bibliography{3DMOTv4}

\end{document}